\documentclass[prd,aps,12pt,nofootinbib,reprint]{revtex4-1} 
\usepackage{mathrsfs,amssymb,amsmath,amsxtra,amsfonts}
\usepackage{graphicx,color}
\usepackage{verbatim}
\usepackage[colorlinks]{hyperref}
\usepackage[utf8]{inputenc}

\begin{document}
\sloppypar \sloppy

\title{Do psi-ontology theorems prove that the wave function is not epistemic?}
\author{I. Schmelzer}
\noaffiliation

\begin{abstract}
As a counterexample to $\psi$-ontology theorems we consider a $\psi$-epistemic interpretation of the wave function in the configuration space representation with a configuration space trajectory defining the ontology. This shows that $\psi$-ontology theorems appear unable to prove that the wave function cannot be interpreted as epistemic.

We identify the criterion used to decide if $\psi$ is epistemic or ontological as the misleading part. Different states having overlaps is only sufficient for being epistemic. In an epistemic interpretation the information which completely identifies the wave function also has an objective base in reality -- the preparation procedure, which has to be really executed to prepare a state.  

Naming the $\lambda \in \Lambda$ ``hidden variables'' and considering them as part of the state of the system we also identify as misleading, because it hides the possibility that they may be visible and external to the system. This leads to further misconceptions about the consequences of the theorems. 
\end{abstract}

\maketitle

\newcommand{\Sch}{Schr\"{o}\-din\-ger}
\newcommand{\Schr}{Schr\"{o}\-din\-ger }

\begin{flushright}
 \emph{But in 1952 I saw the impossible done.} 
 
 J.S. Bell
\end{flushright}

\section{A misleading theorem}
 
Recently, a new type of impossibility theorems has appeared -- the so-called $\psi$-ontology theorems. Their aim is to exclude epistemic interpretations of the wave function by showing that the wave function has to be part of the ontology of quantum theory. Particular examples are \cite{PBR}, \cite{Hardy}, \cite{Gao}, for an extended review of this research, see Leifer \cite{Leifer}. We will mainly refer to \cite{PBR}, but the problem is shared by other variants of the theorem too. What is claimed is that such theorems ``show that any model in which a quantum state represents mere information about an underlying physical state of the system must make predictions which contradict those of quantum theory'', as claimed in \cite{PBR}.

But, similar to Bell in the case of the impossibility theorems of hidden variables, I saw the impossible done, in a $\psi$-epistemic realist interpretation of quantum theory named ``entropic dynamics''. It has been proposed 2011 by Caticha  \cite{Caticha}. 

The question appears what is wrong: Is it the impossibility theorem or the explicit counterexample? As usual (for example in no-hidden-variables theorems),  there is no error, neither in the theorems themselves (which prove what is claimed) nor in the $\psi$-epistemic interpretation.  Simply, if one looks at the  $\psi$-epistemic interpretation, and compares it with the definition, it appears to be, formally, a $\psi$-ontological interpretation too.

So, the theorem is not wrong but misleading, because it relies on a misleading definition of $\psi$-ontology. A definition of $\psi$-ontological interpretations where a clearly $\psi$-epistemic interpretation appears to be $\psi$-ontological should be rejected as inappropriate. 

Behind this, we identify yet another misleading concept -- the idea that a space $\Lambda$ which allows to compute the expectation values of quantum observables $o$ in dependence of choices $a$ of what to measure by the formula 
\begin{equation}\label{basic}
E(o|a) = \int_{\lambda \in \Lambda} o(a,\lambda) \rho(\lambda) d \lambda.                                                                           \end{equation}
somehow has to be a space of ``hidden variables of the quantum system''.  But nothing prevents to use quite visible elements of the external world (or the theoretical discourse about it) as elements of the space $\Lambda$. This is what, essentially, allows an epistemic interpretation of the wave function, where it describes the knowledge about the (objective, real, but external) preparation procedures to define a probability distribution over $\Lambda$. 

\section{A variant of Caticha's ``entropic dynamics''}

Given that Caticha's ``entropic dynamics'' from 2011 \cite{Caticha} has not been cited in the review article of Leifer \cite{Leifer}, it cannot be assumed to be known even by those working in this domain. Thus, for self-consistency, I will start with a short introduction into the basic concepts. In fact, I will present a slightly modified version. 

The goal of Caticha was ``to derive quantum theory as an application of entropic inference''. The interpretation has formal similarities with Nelsonian stochastic mechanics \cite{Nelson}. In particular, it uses the same fundamental ontology, defined by the particle positions in non-relativistic many particle theory. I prefer, instead, to specify the fundamental ontology in a slightly more general way, as defined by some classical configuration space $Q_{sys}$ of the quantum system\footnote{All what one needs for the formalism to work is that the energy depends on the momentum variables in a quadratic form, $H = \langle p,p\rangle + V(q)$. This is fulfilled, in particular, for many relativistic fields, as well as their spatial lattice regularizations.}. 

Caticha's basic assumption is that ``in addition to the particles of interest there exist some extra variables''. ``The number and nature of the extra variables \ldots and the origin of their uncertainty need not be specified''. I will use this freedom to specify them nonetheless, namely as the part $Q_{ext}$ of the configuration space of the whole universe $Q_{univ}$ which is external, outside the system under consideration, so that $Q_{univ}\cong Q_{sys} \times Q_{ext}$. In the language of the Copenhagen interpretation, it would be named the classical part, but in Caticha's approach we do not have to make nontrivial assumptions about this external part, like that it follows classical equations. 

Now, following Bayesian logic, incomplete knowledge about the world would be described by a probability distribution $\rho(q_{sys},q_{ext})d q_{sys} dq_{ext}$ on the space which describes the ontology, that means, on $Q_{sys} \times Q_{ext}$. According to Caticha's interpretation, what defines the wave function of the system is defined by this global probability distribution by the following formulas: First, the resulting probability distribution for the configurations of the system $q_{sys}\in Q_{sys}$, defined by: 
\[\rho(q_{sys}) = \int_{Q_{ext}} \rho(q_{sys},q_{ext}) dq_{ext},\]
together with the entropy of the rest of the world given the system is in a particular configuration $q_{sys}\in Q_{sys}$, defined by 
\[ S(q_{sys}) =  -\int_{Q_{ext}} \ln\rho(q_{sys},q_{ext})\rho(q_{sys},q_{ext}) dq_{ext}. \]
These functions will define the phase of the wave function by
\[\phi(q_{sys}) = S(q_{sys}) - \ln\rho^{\frac12}(q_{sys})\]
and the wave function itself by 
\[ \psi(q_{sys}) = \rho^{\frac12}(q_{sys}) e^{\frac{i}{\hbar}\phi(q_{sys})}.\]

For further details, in particular the derivation of the \Schr equation, we can refer to the original paper \cite{Caticha}, given that it is not relevant here. The known objections are also not relevant: The necessity of a preferred frame is shared with all realistic interpretations because of Bell's theorem, and the Wallstrom objection \cite{Wallstrom,Wallstrom_a} shows only that the interpretation is in fact a different theory, with the additional restriction that in the configuration space $|\psi|^2(q)>0$ everywhere\footnote{The viability depends on the choice of the configuration space. In \cite{Wallstrom2} it is argued that for a field ontology an empirical falsification would become more problematic in comparison with a particle ontology.}. Even if this restriction would not be viable empirically, it would nonetheless define a $\psi$-epistemic realistic interpretation of quantum theory with a \Schr equation, thus, provide a counterexample to the goal of $\psi$-ontology theorems. 

\section{A misleading definition of what is ontological}

So let's take a look at what is wrong with the PBR theorem. For this, it appears sufficient to look at what is proven: 

\begin{quote}
Suppose that, for any pair of distinct quantum states $|\psi_0\rangle$ and $|\psi_1\rangle$, the distributions $\mu_0 (\lambda)$ and $\mu_1 (\lambda)$ do not overlap: then, the quantum state $|\psi\rangle$ can be inferred uniquely from the physical state of the system and hence satisfies the above definition of a physical property. Informally, every detail of the quantum state is 'written into' the real physical state of affairs. But if $\mu_0 (\lambda)$ and $\mu_1 (\lambda)$ overlap for at least one pair of quantum states, then $|\psi\rangle$ can justifiably be regarded as 'mere' information.

Our main result is that for distinct quantum states $|\psi_0\rangle$ and $|\psi_1\rangle$, if the distributions $\mu_0 (\lambda)$ and $\mu_1 (\lambda)$ overlap (more precisely: if $\Delta$, the intersection of their supports, has non-zero measure) then there is a contradiction with the predictions of quantum theory. \cite{PBR}
\end{quote} 

Would Caticha's interpretation fulfill the non-overlap condition, which is used to define $\psi$-ontology? To answer this question, we have to remember what fixes the wave function. According to the minimal interpretation, to prepare a pure state we need some preparation procedure. It is defined by a measurement device measuring some maximal set of commuting operators $\vec{A}$ and a particular choice of eigenvalues $\vec{a}$ of these operators. An unknown initial state of the system has, then, to be measured with the measurement device, and if the measurement gives the result $\vec{a}$, then the system is prepared in the eigenstate $|\psi_{\vec{a}}\rangle$ so that $\vec{A}|\psi_{\vec{a}}\rangle = \vec{a}|\psi_{\vec{a}}\rangle$. This is the information we need, else we would not know that the system is prepared in the particular state $|\psi_{\vec{a}}\rangle$. 

Now, the state of this measurement device is defined by its configuration  $q_{ext}\in Q_{ext}$, which is part of the external world. The measurement results also have to be stored somewhere outside the system, as part of $q_{ext}\in Q_{ext}$. But a particular measurement device which measured $\vec{A}$ and has obtained the particular result  $\vec{a}$ has prepared a state with the wave function $|\psi_{\vec{a}}\rangle$ and no other wave function. So, the real state of the external part $q_{ext}\in Q_{ext}$ defines if a particular state $|\psi_{\vec{a}}\rangle$ has been prepared or not. 

Can there be any overlap between the $q_{ext}$ preparing eigenstates $|\psi_{\vec{a}}\rangle$ of $A$ and those preparing eigenstates $|\psi_{\vec{b}}\rangle$ of some other $B$? Certainly not. There may be many different preparation procedures $q_{ext}$ all preparing the same state $|\psi_{\vec{a}}\rangle$ of $A$. But the preparation procedures for different operators differ, and different eigenstates prepared by the same preparation measurement differ nonetheless in the results of this initial measurement. So, there can be no overlap, and therefore, if we follow the consideration quoted above, ``every detail of the quantum state is 'written into' the real physical state of affairs''.

So, it is not the theorem itself which is the problem. What the theorem proves -- that an overlap would lead to a contradiction with quantum predictions -- is unproblematic because there is none in our $\psi$-epistemic interpretation. 

\section{State of the system or state of the world?}

Let's note that there is a difference in the interpretation of the meaning of ``hidden variables'':

\begin{quote}
The result is in the same spirit as Bell’s theorem [.], which states that no local theory can reproduce the predictions of quantum theory. Both theorems need to assume that a system has a objective physical state $\lambda$ such that probabilities for measurement outcomes depend only on $\lambda$.  \cite{PBR}
\end{quote}
In the $\psi$-epistemic interpretation considered above this is not the case -- the information which defines the wave function is not part of the physical state of the system, but part of the world outside of the system -- it is contained in the preparation device and the result of the preparation measurement.

Thus, the ``hidden variables'' $\lambda$ of the ``state of the system'' are in the epistemic interpretation neither part of the system nor hidden. Instead, they simply contain the ``incomplete knowledge'' which defines the wave function, because it is part of the complete ontology of the universe. There is nothing in assumptions of the theorem which would force the variables to be hidden or to be defined by the system taken alone instead of the whole world. So, it is only this implicit assumption which is the problem, and not that the epistemic interpretation does not fulfill this assumption. 

\section{Misleading consequences}

This confusion is not innocent, because it suggests consequences which are not justified.  

The wave function being something defined by hidden variables of the system itself would require some sort of additional degrees of freedom beyond the configuration space trajectory of the system, variables able to store quite arbitrary wave functions. While it does not follow in any strong sense, what seems natural would be some sort of storage sufficient to describe arbitrary wave functions. The epistemic interpretation above shows that there is no necessity for any such additional storage inside the system. 

Another implicit assumption would be that this storage would have to be very large, able to store arbitrary wave functions. To quote PBR \cite{PBR}:
\begin{quote}
\ldots the quantum state has the striking property that the number of real parameters needed to specify it is exponential in the number of systems n. This is to be expected if the quantum state represents information but is -- to us -- very surprising if it has a direct image in reality.
\end{quote}
The phrase ``has a direct image in reality'' suggests a connection which does not exist, namely a map from the space of all possible wave functions into the space $\Lambda$ which defines the ontology of the theory. If one looks at what we really have in the epistemic interpretation, we find a map from particular preparation procedures (that means, from some subset of possible real states of the external world where pure states of the system have been prepared $q_{prep}\in Q_{prep} \subset Q_{ext}$ into the space of wave functions of the system $\psi(q_{sys}) \in \mathcal{L}^2(Q_{sys})$. 

It follows that the set of possible wave functions is restricted by the set of possible configurations of the external world, a world of the same type as the quantum system itself, defined by a single configuration $q_{ext}\in  Q_{ext}$ instead of a wave function $\psi(q_{sys}) \in \mathcal{L}^2(Q_{sys})$. 

In other words, there is no image of a wave function in reality. There may be only a pre-image. And it is quite plausible that the existence of such a pre-image will be a rare exception instead of the rule.

\section{Discussion}

We have found that the definition of what defines a $\psi$-ontological theory is seriously misleading, because it classifies a clearly $\psi$-epistemic interpretation of \Schr theory as $\psi$-ontological. 

As the confusion behind this misleading definition we have identified the very notion of ``hidden variables of a system'' $\lambda \in \Lambda$ which define measurement results $o$ in dependence of decisions of the experimenters $a$ by 
 
It is somehow presumed that this space $\Lambda$ defines a space of elements which are, first of all, hidden, and, moreover, part of the quantum system. But in the case of the epistemic interpretation we have used as a counterexample, the space $\Lambda$ contains, together with the configuration of the system, also the configuration of the external world, thus, it is neither hidden nor part of the system.  

In the proofs using this formula nothing depends on these implicit assumptions. So, they can be applied also if the space $\Lambda$ is a visible part of the external world. As long as this is not recognized, it has misleading consequences. 

In the case of Bell's theorem, the misleading part is that one can somehow get rid of the formula by rejecting ``realism'' which requires it for some space of ``hidden variables''. One cannot, because the space $\Lambda$ can be constructed in the objective Bayesian probability interpretation -- the ``logic of plausible reasoning'', simply starting with all the propositions under consideration the particular field of discourse, namely as the set of all ``logically imaginable possibilities'' -- logically consistent assignments of truth values to all propositions \cite{Belllogic}. So, the rejection-of-realism loophole does not exist, those who believe in it are misguided. 

The fate of the $\psi$-ontology theorems is similar. They implicitly depend on the assumption that something nontrivial about the ``hidden variables of the system'' has been proven. But the space of ``hidden variables of the system'' can simply include the whole preparation procedure, which is visible and external. Once this is understood, it becomes clear that a $\psi$-epistemic interpretation, where the implicit knowledge about a pure state is based on the knowledge of the objective preparation procedure used to prepare it, may become a $\psi$-ontological interpretation if one follows the definition -- simply because the ``hidden variables of the system'' contain that visible preparation procedure located in the external world too. Those who believe that it follows from the theorems that the wave function should be interpreted as part of the ontology are misguided. 

To summarize, the $\psi$-ontology theorems are unable to show that the wave function is not purely epistemic, and misleading by suggesting this. This has been shown explicitly by the presentation of a clearly $\psi$-epistemic interpretation of the wave function based on an ontology completely described by a single configuration space trajectory.  


\end{document}